\documentstyle[aps,prb,epsfig,floats]{revtex}
\begin{document}

\draft
\wideabs{

\title{Quantum vortex fluctuations in cuprate superconductors}

\author{Hyok-Jon Kwon }
\address{Department of Physics and Center for Superconductivity
Research, University of Maryland, College Park, MD 20742-4111}
\date{October 9, 2000}

\maketitle
\begin{abstract}

We study the effects of quantum vortex fluctuations in two-dimensional
superconductors using a dual theory of vortices, and investigate the
relevance to underdoped cuprates where the superconductor-insulator
transition (SIT) is possibly driven by quantum vortex proliferation. We
find that a broad enough phase fluctuation regime may exist for
experimental observation of the quantum vortex fluctuations near SIT in
underdoped cuprates. We propose that this scenario can be tested via
pair-tunneling experiments which measure the characteristic resonances in
the zero-temperature pair-field susceptibility in the vortex-proliferated
insulating phase.

 \end{abstract}
}

\section{Introduction}

The emergence of the pseudogap phase in  cuprate superconducting materials
may be attributed to  strong order parameter phase fluctuations.
Based on the small superfluid phase stiffness (SPS) at $T=0$, together with its
empirical scaling with the transition temperature
($T_{\rm c}$), a  proposal was put
forward
 that the normal state pseudogap phase in the underdoped regime is
a superconductor whose phase coherence is destroyed by  thermal
phase fluctuations but with a robust gap-like feature due to a
strong pairing amplitude\cite{Uemura,pseudogap,ours}. 
Some experimental evidence was provided
for strong thermal fluctuations of unbound vortices over a wide range
of temperatures above $T_{\rm c}$ in the
pseudogap phase\cite{Corson}. This suggests that the superconducting
transition is of  Kosterlitz-Thouless type
 with a broad phase fluctuation regime.

It is then natural to explore the effects of  strong $T=0$ quantum phase
 fluctuations, since the strengths of both  thermal and
 quantum phase fluctuations are
correlated  with the small magnitude of  SPS.
For instance, quantum phase fluctuations reduce the Debye-Waller
factor $\langle e^{i\phi}\rangle$ where $\phi$ is the order  parameter
phase, and have significant consequences for the $c$-axis optical
conductivity\cite{Ioffe}, the renormalization of  SPS, and the  pair-field
susceptibility\cite{Paramekanti}. 
A more dramatic effect of  quantum phase fluctuations may be
the quenching of  phase-coherence by vortex pair proliferation 
in the underdoped regime, which leads to a superconductor-insulator
transition (SIT)\cite{SIT1,SIT2}. 
Some experimental findings are consistent with the
existence of a quantum critical point  near SIT\cite{QCP,Decca}, controlled
 by the charge carrier doping. 
However, direct evidence for  vortex
pair proliferation is yet to be found, although an experiment on
current-voltage ($I$--$V$) characteristic in Bi2212 shows indications
 of  a large density of  quantum vortex pairs well below $T_{\rm c}$ and far
away from SIT\cite{Chiaverini}.
When searching for definitive experimental tests, it is important to
 construct and study reliable 
phenomenological theories which incorporate  vortex fluctuations.
The critical properties of vortex-proliferated SIT
have been studied using a  framework of the dual
transformation. In this framework, dual
vortex fields are conveniently introduced as the
new order parameter of vortex proliferation\cite{SIT1,dual,soliton},
and the transition is described by the 2+1-dimensional (2+1D)
 Ginzburg-Landau theory of
the quantum vortex order parameter, although the actual critical
properties are determined by the existence of long-range
interactions, periodic potential, and disorder\cite{Boson}.
In this paper, we directly apply the mean-field 
dual formulation  to two-dimensional (2D) superconductors and 
study its phenomenological relevance to  cuprates in both
superconducting and  insulating states.  We find that the 
phase fluctuation regime may be broad enough for experimental access. 
 In vortex-proliferated insulating states,
the characteristic form of the pair-correlation function is
$\langle e^{i\phi({\bf x},t)}e^{i\phi(0,0)}\rangle
\propto e^{ia\sqrt{|t|^2-|{\bf x}|^2/v_{\rm v}^2}}$, with $a>0$.
We propose that this novel form of the pair-correlation function can be
verified by pair-tunneling experiments.

\section{Dual theory of quantum vortices}
In this paper, we denote the space-time 3-vectors with $x$ and $k$ and
the spatial vectors with ${\bf x}$ and  ${\bf k}$. We use Greek
indices running from 0 to 2 and Roman indices $i,j$ from 1 to 2.
We also adopt a  metric tensor  
$g^{\mu\nu} =\delta^{\mu\nu}(1-2\delta ^{\mu 0})$ to evaluate
contractions of indices.  For notational convenience, we set $\hbar
=1$. We begin with a $T=0$ BCS model of a  2D
superconductor coupled to the electromagnetic field.
\begin{eqnarray}
& &S = \int dt \int d^2x\sum_\sigma \bigg{[}
c^{\dag}_\sigma \left(iD_t -{D_{\bf x}^2 \over 2m} -\mu\right)
 c_\sigma   \nonumber \\ \nonumber
&&\quad +\Delta ({\bf x},t)
c^{\dag}_\uparrow({\bf x},t) c^{\dag}_\downarrow({\bf x},t)
 + {\rm h.c.} +{1\over g}|\Delta({\bf x},t)|^2 \bigg{]}+  S[A_\mu]~,
\end{eqnarray}
where $c_\sigma$ is a fermion field,
$D_\mu = \partial_\mu -iA_\mu$ is the covariant derivative,
$A_\mu$ is the electromagnetic vector potential, and $m$ is the
effective fermion mass.
Here we do not specify the pairing symmetry, since
it is not significant within the accuracy of our discussion.
$S[A_\mu]$ is an electromagnetic gauge field action which takes the
following form in 2D:
\begin{eqnarray}
S[A_\mu]=\sum_{k}\bigg{[} {|{\bf k}|\over 4\pi e^2}A_0^2 
-{({\bf k}^2c^2-\omega^2)d\over 8\pi e^2}{\bf A}^2\bigg{]}~,\nonumber
\end{eqnarray}
where we chose the Coulomb gauge $\nabla\cdot {\bf A}=0$ and $d$ is
the thickness of the film.
We assume that the order parameter amplitude fluctuations    
are negligible [$|\Delta ({\bf
x},t)|=\Delta$], and focus on the 
phase degree of freedom $\phi =-i\ln [ \Delta ({\bf x},t)/\Delta ] $.
In order to decouple the
$\phi  $ field from the order parameter amplitude, we perform a
singular gauge transformation
$ \psi_\sigma({\bf x},t) = c_\sigma({\bf x},t)
e^{-i\phi ({\bf x},t)/2}$, with $\psi_\sigma$ the field operators for
the transformed quasiparticle\cite{comment1}.
We eventually arrive at the following
 effective theory of the phase and the electromagnetic
$A_\mu$ fields after integrating out the fermion degrees of 
freedom\cite{Paramekanti}:
\begin{eqnarray}
&&S[\phi,A_\mu] ={1\over 2} \int d^2x~dt \left\{ 
\rho_{\rm s}\left[ (2A_0+\partial_t \phi)^2/c_s^2 \right. \right.\nonumber \\
&&\left. \left. -(2{\bf A}+\nabla\phi
)^2\right] +\partial_t \phi ~n_s \right\} +S[A_\mu]~,
\label{phiA}
\end{eqnarray}
where $\rho_{\rm s}$ is the 
SPS defined as $\rho_{\rm s} = n_s/4m$,  $n_s$ is the superconducting
fermion density,   and
$c_s $ is an analog of the phonon velocity in a superfluid. 
In a 2D BCS superconductor, $c_s $ is related to the Fermi velocity $v_F$
by $c_s=v_F/\sqrt{2} $.   We are interested in deriving the
effective theory of the phase fields which describes  vortices. Therefore,
we separate $\partial _\mu \phi$ into $\partial _\mu \phi =\partial
_\mu  \theta +{\cal A}_\mu$ where $\theta $ is the spin-wave type
Goldstone fluctuation and ${\cal A}_\mu$ gives
topologically non-trivial phase gradients generated by vortices. Then
we integrate 
out both $\theta$  and  $A_\mu$ fields to obtain the 
final effective action of ${\cal A}_\mu$:
\begin{eqnarray}
&&S[{\cal A}_\mu]={1\over 2}\sum_{k}[ {\cal K}_0(k)|{\cal A}_0|^2
-{\cal K}_T(k)|{\vec{\cal  A}}|^2 ]~,
\label{SA}\label{SJ}
\end{eqnarray}
with a constraint that $\nabla \cdot \vec{\cal A} =0$.
Here
${\cal K}_0 = \rho_{\rm s} {\bf k}^2/({\bf k}^2c_s^2- \omega ^2+8\pi
e^2\rho_{\rm s} |{\bf
k}|)$
and 
${\cal K}_T =\rho_{\rm s} (-\omega^2/c^2 +{\bf k}^2)/(\lambda^{-2}+{\bf
k}^2-\omega^2/c^2)$ where $\lambda $ is the penetration depth. 
Over the time scale of our interest, ${\cal K}_T\approx \rho_{\rm s} {\bf
k}^2/(\lambda^{-2}+{\bf k}^2)$.  Below we will neglect the term
$\partial_0 \phi~n_s/2$ in Eq. (\ref{phiA}), which plays the role of a
dual magnetic field but is insignificant near SIT since $n_s$ is
renormalized and approaches zero in the insulating state. 

Now we assume a certain distribution of $ N$ point-like vortices. 
At a fixed time, the vortices in 2D superconductors are
 pancake-like. In 2+1D, however, we can consider the space-time paths of
pancake vortices as 3D lines of vortices.
The current density (${\cal J}_\mu$) of the vortices are obtained from 
 ${\cal J}^\mu =\epsilon^{\mu \nu \lambda}
\partial_\nu {\cal A}_\lambda$.
${\cal J}_\mu$ can be expressed in term of the  space-time
paths of vortices in 2+1D as follows:
\begin{equation}
{\cal J}^\mu (x) =2\pi \sum _{l=1}^N\int du_l~{d\bar{X}_l^\mu (u_l)\over
du_l}\delta^{(3)}(\bar{X}_l(u_l)-x)~,
\label{Jp}
\end{equation}
where $u_l$ parameterizes the  space-time
path $\bar{X_l}$ of the $l$-th vortex. 
Therefore, 
Eq. (\ref{SA}) describes the interactions between infinitesimal vortex
segments in 2+1D (Ref.\cite{QED}) if we re-express $S[{\cal A}_\mu]$ in
terms of  ${\cal J}_\mu$. 
Here we assume only  vortices  of one flux quantum; the ones traveling
backward (forward) in time are antivortices (vortices). 

The field theory of Eqs. (\ref{SA}) and (\ref{Jp}) is
 inconvenient for  description of particle-like vortices. Therefore,
 we first transform   Eqs. (\ref{SA}) and (\ref{Jp}) into  particle dynamics.
 Later  we will conveniently  transform the particle dynamics 
 into field dynamics of vortices. Before the transformation, we
first separate the vortex-current interactions in Eq. (\ref{SJ}) into 
contact (short-range) and 
long-range interactions. The contact interaction can be
approximately expressed in the form of a relativistic particle action
as follows:
\begin{eqnarray}
&&S_{\rm cont} \approx \sum_l \int du_l~m_{\rm v}\left[ - v_{\rm v}^2\left(
{d\bar{X}^0_l\over du_l}\right) ^2+{1\over 2}\left(
{d\bar{X}^i_l\over du_l}\right) ^2 \right],
\label{local}
\end{eqnarray}
where we have deduced the rest mass  from the
contact interaction as following:
\begin{eqnarray}
m_{\rm v} v_{\rm v}^2 ~\delta^{(0)}(0)&=& (2\pi)^2\sum_k {\cal
K}_T(k)/2{\bf k}^2~. 
\label{mass}
\end{eqnarray}
 The velocity $v_{\rm v}$ is taken as
$v_{\rm v}= Cv_F$ where $C =O(1)$ on physical grounds\cite{QED}. 
In addition to Eq. (\ref{local}), there is
 contact repulsion between  vortices of distinct labels.
We will not estimate or discuss this repulsive interaction
explicitly, except to mention
that it stabilizes the non-zero expectation value   of vortex fields 
in the insulating state which we will
discuss later.  
In order to incorporate the
long-range interactions into the particle dynamics, we first 
separate the dynamics of ${\cal J}_\mu$ and ${\cal A}_\mu$ by introducing a
Lagrange multiplier $G_\mu$. This amounts to adding a term $\int d^2x~dt~G_\mu(
\epsilon^{\mu\nu\lambda} \partial_\nu{\cal A}_\lambda - {\cal J}^\mu)$
to the action in Eq. (\ref{SA}) which
enforces the relation between ${\cal J}_\mu$ and ${\cal A}_\mu$. 
 Then we integrate out  
${\cal A}_\mu$ fields to obtain the following effective interactions:
\begin{eqnarray}
&&S_{\rm int} ={1\over 2} \sum_{k}\left[ 
{{\bf k}^2 +\lambda^{-2}\over \rho_{\rm s} }
|G_0|^2 
-{({\bf k}^2c_s^2+8\pi e^2\rho_{\rm s}|{\bf k}|-\omega^2)\over
\rho_{\rm s}}
|{\bf G}|^2 \right] \nonumber \\
&&\quad +2\pi \sum_l \int du_l~
G_\mu(\bar{X}_l)~d\bar{X}_l^\mu/ du_l~, \label{non-l}
\end{eqnarray}
with a Coulomb gauge constraint $\nabla \cdot {\bf G}=0$. 
Thus, combining Eqs. (\ref{local}) and (\ref{non-l}), we have obtained a
theory of relativistic particles with a rest mass $m_{\rm v}$,
 coupled to the field $G_\mu$ which mediates the superfluid phase
 modes between the vortices.

We are now in a position to transform the particle dynamics into a
field theory. Since we are interested in vortex-antivortex pair
creation (annihilation), we now specialize into distribution of 
 2+1D  vortex loops  by requiring that the paths
$\bar{X}(u)$ are closed trajectories.
Here we follow Ref. \cite{soliton} and use the particle-field
correspondence to re-express
the total action $S_{\rm cont}+S_{\rm int}$  
in terms of a  relativistic complex scalar field $\Phi$
coupled to the gauge field $G_\mu$:
\begin{eqnarray}
&&S[\Phi,\Phi^*,G]=\int d^2x~dt~ [ -(\partial_\mu-2\pi iG_\mu)\Phi^*
 (\partial^\mu+ 2\pi i G^\mu)\Phi\nonumber \\
&&\quad -m_{\rm v}^2|\Phi|^2]+S[G_\mu]~,
\label{Sd}
\end{eqnarray}
where $G_\mu $ represents the local U(1) gauge symmetry of $\Phi$ fields.
Here $S[G_\mu] $ is the part quadratic in $G_\mu$ from
Eq. (\ref{non-l}) and we set $v_{\rm v} =1$ for notational 
convenience. This is the well-known dual form of the theory of 
superconductivity where the roles of magnetic (vortices) and
electric (Cooper pairs) charges are interchanged\cite{dual}.
It is easy to show that the vortex current can be related to $\Phi$ as
${\cal J}_\mu/2\pi = i(\Phi\partial_\mu \Phi^*-\Phi^*\partial_\mu
\Phi)+4\pi G_\mu|\Phi|^2$, which automatically satisfies the necessary
conservation condition $\partial_\mu {\cal J}^\mu =0$.
 
\section{Strength of quantum vortex fluctuations in cuprates}
Next,  we wish to  explore the feasibility of
experimental observation by 
studying the width of the phase fluctuation regime where we expect to observe
strong phase fluctuation effects. We assume a linear scaling of SPS with
charge-carrier doping, and search for the magnitude of SPS (doping) at
the most likely point of SIT  in the underdoped cuprates.
 The action in Eq. (\ref{Sd}) can be viewed as the effective
3D Ginzburg-Landau  functional of $\Phi$. 
Then  the SIT  occurs upon ordering [U(1)-symmetry breaking]
 of the $\Phi$ fields
 when $m_{\rm
v}^2<0$, so that $\langle \Phi \rangle \neq 0$ 
In this case,
 $\langle \Phi(x) \Phi^*(y)\rangle \approx \Phi_0^2 $, where
$\Phi_0^2$ is a positive constant in the mean-field approximation,
independent of $|x-y|$. 
The correlation function $\langle \Phi(x)\Phi ^*(y) \rangle$ can be
considered as the expectation value of the number of vortex paths that 
connect $x$ and $y$. Therefore, $\langle \Phi(x)\Phi ^*(y) \rangle $=
constant implies a non-zero and constant probability
of arbitrarily long 2+1D vortex loops.
The non-zero expectation value of $\Phi$  leads to
 recovery of the U(1)-symmetry of the superconducting order parameter,
and hence the Meissner effect is absent even in the presence of a
non-zero bare SPS. Below we  discuss the renormalization of 
$m_{\rm v}^2$ to one loop expansion, assuming that the loop expansions
are reliable.

\begin{figure}[t]
\centerline{\psfig{file=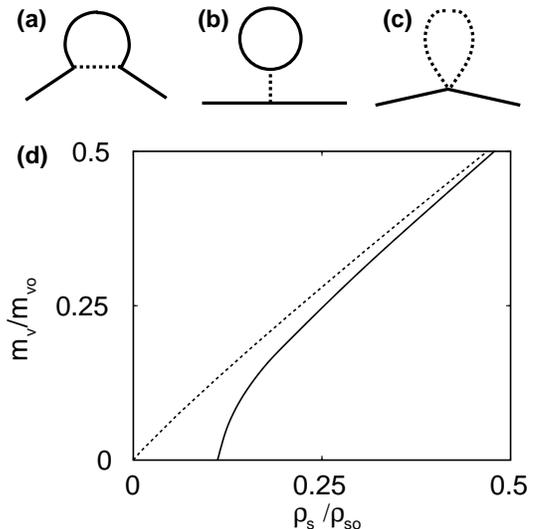,width=0.8\linewidth,angle=0}}
\caption{(a)--(c) Diagrams of one-loop corrections to $m_{\rm v}^2$. The solid
(dashed) lines denote  $\Phi$ ($G_\mu$) field propagators.
(d) Solid (dashed) line is the one-loop-renormalized (bare) vortex
rest mass ($m_{\rm v}$) as a function of the bare superfluid phase
stiffness ($\rho_{\rm s} =n_s/4m$). 
$m_{\rm vo}$ and $\rho_{\rm so}$ are respectively the vortex rest mass
and  the superfluid phase stiffness at optimal doping.
}
\label{fig:loop}
\end{figure}

 We use  following parameters of optimally doped
cuprates:  $\Delta \approx$ 20 meV, $\lambda \approx 200~{\rm
nm}$, the coherence length 
$\xi \approx$ 2 nm, and the film thickness $d\approx$ 1.5 nm which roughly
corresponds to  a monolayer film.  We assume a vortex core size
of $\xi$ which provides a momentum (short-range) cutoff at
$\Lambda_c=\pi/\xi$  and a
frequency (short-time) cutoff at $v_F\Lambda_c$. With this prescription, 
we avoid ultraviolet divergences and obtain 
$ \delta^{(3)}(0)=v_F\Lambda_c^3/(2\pi)^3$. 
We can then estimate $m_{\rm v}$ and $v_{\rm v}$ from Eq. (\ref{mass}) in
terms of the above parameters. 
The result is $m_{\rm v}v_{\rm v}^2 \approx \rho_{\rm s}\pi\ln \kappa $ 
where $\kappa = \lambda\Lambda_c $.
We find that $m_{\rm v} \approx 0.36 $ eV 
in optimally doped cuprates.
For simplicity, we assume that  
the  charge-carrier doping only affects  magnitudes of
the bare SPS, and we keep the other parameters constant.
 From Eq. (\ref{Sd}) we can calculate  one-loop
corrections to $m_{\rm v}^2$ as shown in Fig. \ref{fig:loop}
and find the point where the correction 
is of the same order of magnitudes as  the bare value.
In fact, the only correction comes from Fig. \ref{fig:loop}(a);
Fig. \ref{fig:loop}(b) does not contribute because the $\Phi$ loop
vanishes, and  
 Fig. \ref{fig:loop}(c)  has been
already included 
as the contact interaction. The correction $\delta m_{\rm
v}^2/m_{\rm v}^2$ roughly behaves as 
$-8(v_{\rm v}\Lambda_c)^3/3\rho_{\rm s}^3\pi^4\ln^3\kappa $, 
whose magnitude becomes  large when the bare
SPS is smaller.
Eventually, we find  that 
$m_{\rm v}^2-\delta m_{\rm v}^2 \rightarrow 0$ (breakdown of
Ginzburg-Landau theory)
 occurs when  $\rho_{\rm s} \approx 0.3\Delta/\ln\kappa \approx 2$ meV for cuprate  
films, which corresponds to about 10 \% of that of the optimal doping 
[See Fig. \ref{fig:loop}(d)]. This implies that, using the empirical
scaling  between $\rho_{\rm s} $ and $T_{\rm c}$, the phase fluctuation regime
begins when $T_{\rm c}$ is less than 10 K or so. In terms of charge-carrier
(hole) doping concentration $p$, 
this corresponds to  $p\approx 0.35 p_{\rm o}$ where $p_{\rm o}$ is the optimal
doping concentration. This is 
from the empirical relation between the critical temperature and
doping\cite{Deutscher}, $T_{\rm c}/T_{\rm co} \approx 1-2.1(p/p_{\rm o}-1)^2$, where
$T_{co}$ is the optimal critical temperature and $T_{\rm c}\approx 0$ for
$p\approx 0.3 p_{\rm o}$.
Therefore, there may be a broad enough doping range ($0.3 p_{\rm o}
<p< 0.35 p_{\rm o}$) 
 for experimental studies on critical vortex fluctuations such as non-linear 
$I$--$V$ characteristic\cite{Chiaverini}. For instance, the underdoped
YBCO used in the anomalous proximity effect experiment performed by
Decca {\it et al.}\cite{Decca} fall within this parameter regime,
which indicates strong phase fluctuations.  
The critical properties of the SIT are beyond the scope of
this paper. Instead, we will simply assume  a non-zero expectation
value of $\Phi$
and discuss  definitive experimental tests of  vortex-proliferated
 insulating states.

\section{Pair-field susceptibility}
In order to establish that underdoped cuprates are under
strong quantum vortex fluctuations, a  direct probe into the
pair-fluctuations is necessary. Here we propose a pair-tunneling
experiment\cite{Carlson} 
to measure the pair-field susceptibility which contains
information about   pair-correlation functions. We first give
a brief summary of the desired experimental setup.
We consider a $c$-axis tunnel junction of a thickness $\delta$ 
between two cuprate
samples, where one of them is   optimally doped with a
$c$-axis penetration depth $\lambda_c$, and the
other is an underdoped insulating film at $T=0$ with a thickness $d$.
Since one of the electrodes is insulating, the usual Josephson current
oscillating at a frequency of 
$2eV/\hbar$ is absent.
However,
an  {\it excess} current will flow due to Josephson coupling of the 
superconducting pair-field of the superconducting electrode to the 
{\it fluctuating} pair-field of the insulator.  
Neglecting vortex fluctuations in the optimally doped electrode, 
the excess current 
can be related to the pair-field susceptibility of the insulator as
$I_{\rm ex}=(eE_J^2/S\hbar^2) {\rm Im}~ D^R(q,\omega)$ where
$D^R({\bf x},t)=-i\theta(t)\langle [e^{i\phi({\bf x}t)},e^{i\phi({\bf
0}0)}] \rangle $ is the retarded pair-correlation function,
 $E_J$ is the Josephson coupling energy of the junction, and $S$ is
 the junction contact area. Here $\omega$ is a frequency
$2eV/\hbar$  and $q$ is a wave vector $q(H)= 2eH(\lambda_c+d/2+\delta)/\hbar
c$ which is determined by a small magnetic field $H$  applied
parallel to the junction \cite{Scal}. Therefore, the excess current
 can provide information about the spectrum of phase fluctuations.
 Below we obtain the form of the
 pair-correlation function of  vortex-proliferated insulators.

\begin{figure}[t]
\centerline{\psfig{file=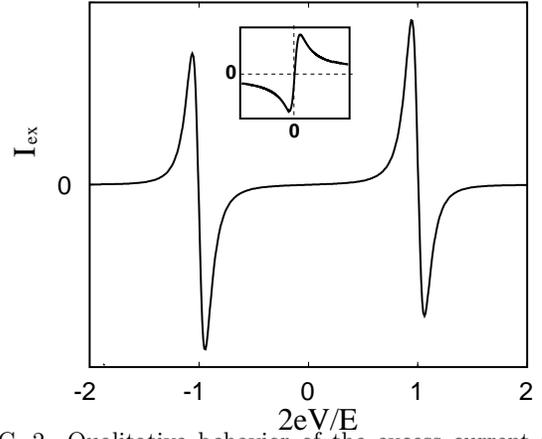,width=0.8\linewidth,angle=0}}
\caption{Qualitative behavior of the excess current in an arbitrary
unit. For comparison, the excess current due to the
conventional finite-temperature fluctuations is shown in the inset.}
\label{fig:Exc}
\end{figure}

First, we re-express the pair-correlation function as
$D({\bf x},t)
= -i\langle \exp[ i\int_{{\bf 0},0}^{{\bf x},t}{\cal A}_\mu(\bar{x})
d\bar{x}^\mu]\rangle$,
ignoring  contributions from $\theta$ fields since they are
sub-leading in  long length scales. Here we assume that 
 $2eV < \Delta$ to avoid  the effect of order
parameter amplitude fluctuations.
Then we introduce a source function $j_\mu (\bar{x}) =\int_{{\bf
0},0}^{{\bf x} ,t}dy_\mu \delta^{(3)}(y-\bar{x})$ where we take $y$ to
be a straight line which connects $({\bf 0},0)$ and $({\bf x},t)$.
Now we can re-express $D$ as following:
\begin{eqnarray}
&&iD({\bf x},t) =
{\int {\cal D}\Phi {\cal D}\Phi^* {\cal D}G_\mu ~ e^{i\int d^3x {\cal
A}_\mu j^\mu +iS[\Phi, \Phi^*, G_\mu]}\over
\int {\cal D}\Phi {\cal D}\Phi^* {\cal D}G_\mu ~e^{ iS[\Phi, \Phi^*,
G_\mu]}}~
\label{funct}\\
&&\quad  \approx e^{ -\sum_{k,\mu} j_\mu(k)j_\mu(-k)\langle {\cal
J}_\mu(k) {\cal J}_\mu(-k)\rangle /2{\bf k}^2 }~, \nonumber
\end{eqnarray}
where the second line is obtained by the method of functional
integration (see Appendix \ref{sec:App}). 
In order to obtain the vortex-current correlation functions, we first
define the polarization functions:
\begin{eqnarray}
\Pi_\mu(x-y)&& = \langle |\Phi|^2\delta^{(3)}(x-y)  -
{\cal P}_\mu(x){\cal P}_\mu^*(y)/2\rangle
\nonumber .
\end{eqnarray}
where ${\cal P}_\mu(x) =
\Phi(x)\partial_\mu\Phi^*(x)-\Phi^*(x)\partial_\mu\Phi(x)$.
We also define the longitudinal ($\Pi_0$) and transverse ($\Pi_T$) polarization
functions  as the $\mu$-component of the polarization functions where
$\mu$ is in the longitudinal ($0$) and transverse ($T$) direction respectively.
Then the vortex-current correlation functions can be written
as following
\begin{eqnarray}
i\langle {\cal J}_0(k){\cal J}_0(-k)\rangle
&\approx &{8\pi^2 {\bf k}^2\Pi_0(k)\over [{\bf 
k}^2+ 8\pi^2{\cal K}_T(k)\Pi_0(k)]}, \nonumber \\
i\langle \vec{\cal J}(k)\cdot \vec{\cal J}(-k)\rangle
&\approx &{-8\pi^2 {\bf k}^2\Pi_T(k)\over [{\bf 
k}^2+ 8\pi^2{\cal K}_0(k)\Pi_T(k)]}~,
\label{JJ}
\end{eqnarray}
Using Eq. (\ref{JJ}) and assuming $\Pi_\mu (k) \approx $ constant $ >0$,
which holds in the vortex proliferated state,
we obtain the asymptotic behavior of $D({\bf x},t)$
in the large $|{\bf x}|$, $|t|$:
\begin{equation}
 iD({\bf x},t) \propto \exp\left[2\pi i\ln(L/\xi)
 \Pi_0\sqrt{ v_{\rm v}^2|t|^2-|{\bf x}|^2}   \right]~,
\label{Pair}
\end{equation}
where $L$ is the size of the sample. 
Upon Fourier transformation, we find that 
$D({\bf q},\omega)\propto \Pi_0/(\omega^2-E^2)^2$ where
$E^2=[2\pi \ln(L/\xi)\Pi_T] ^2+v_{\rm v}^2q^2(H)$.
Accordingly, the excess current behaves as
$I_{\rm ex}({\bf q}, 2eV)\propto 1/[(2eV)^2-E^2]^3$, resembling
the imaginary part of  resonance peaks located near $2eV=\pm 2\pi
\ln(L/\xi)\Pi_T$. The resonance peak heights are determined by
 the normal-state junction resistance and $\langle |\Phi |^2 \rangle$. 
The apparent logarithmic divergence of the exponent in
Eq. (\ref{Pair}) and of the
resonance energy is due to the fact that the energy of dual vortices
(Cooper pairs) is logarithmically divergent as the system size, similar
to vortices in superfluid helium. This weak divergence does not pose a
serious problem in realistic samples which have finite sizes, however.
This excess current qualitatively differs from that due to 
fluctuations at finite-temperature superconducting
transitions\cite{Carlson} where $I_{\rm ex}\propto \omega/(\omega^2+\Gamma^2)$ 
(See Fig. \ref{fig:Exc}).
Therefore, it is possible to confirm the existence of 
phase-fluctuation driven insulating states by 
detecting   characteristic 
resonances in the pair-field susceptibility. 
The pair-correlation function oscillates 
 at large $|t|$  due to the presence of  vortex
condensates. In the phase-coherent state of the vortex fields $\Phi$, the
trajectories of Cooper pairs act as  {\it dual} vortex paths in
2+1D,  and the pair-correlation  function is the probability of a  dual
vortex\cite{dual} of length  $\sqrt{v^2_{\rm v}|t|^2-|{\bf x}|^2} $. 
Therefore $D({\bf x},t)$ is determined by the
action of a vortex line which connects $({\bf 0},0)$
and $({\bf x},t)$. This explains the time-like length-dependence of
the exponent of $D({\bf x},t)$. 
In fact, due to  dissipative processes that we have not considered,
 the pair-correlation function decreases in magnitude at large
 $|t|$  in addition to the pure oscillation in $|t|$
 shown in Eq. (\ref{Pair}). Accordingly, the resonance  peaks in the
pair-correlation function will be broadened depending
on the strength of the dissipation.

\section{Summary and conclusions}
We discussed the possibility of observing quantum vortex fluctuations
in underdoped cuprate superconductors near SIT.
Using the dual theory of vortices,
we showed that cuprate superconductors are subject to strong
vortex fluctuations so that it is possible to experimentally access
the fluctuation regime near  SIT. As a definitive test
of the phase fluctuation scenario, we proposed  an experiment to measure  the
pair-field susceptibility to probe the form of the pair-correlation
function in the insulating regime.
We expect  that the pair-field susceptibility in
vortex-proliferated insulating states is qualitatively different from
that observed in the normal-state fluctuation regime, and
shows characteristic resonance peaks. 
The result can be generalized to any phase-fluctuation driven SIT of
superconducting films.
We anticipate that  a more realistic dual theory of vortex fluctuations in
 superconducting films can be obtained by including
the effects of  normal fluids or  $d$-wave nodal quasiparticles.

\section{acknowledgement}
The author gratefully acknowledges stimulating discussions with
 A. T. Dorsey,   H. D. Drew, C. J. Lobb, K. Sengupta, A. Sudb\o, and
V. M. Yakovenko. 
This work was supported by the 
NSF DMR-9815094 and the Packard Foundation.

\appendix

\section{Pair-correlation functions}
\label{sec:App}
Here we give a brief derivation of Eq. (\ref{funct}).
From the first line of Eq. (\ref{funct}), the first term in the
 exponent  can be rewritten as
\begin{eqnarray}
\int d^3x {\cal A}_\mu j^\mu = \int d^3x {\cal J}^\lambda h_\lambda \nonumber
\end{eqnarray}
where $h_\lambda$ is defined through the relation 
$j^\mu = i\epsilon^{\mu\nu\lambda}\partial_\nu h_\lambda$.
Assuming a Gaussian functional integral, we use the Baker-Haussdorf
formula to obtain
\begin{eqnarray}
\langle e^{i\int d^3x {\cal J}^\lambda h_\lambda}\rangle \nonumber
\approx e^{i~S_C}~.
\end{eqnarray}
where 
\begin{equation}
S_C = i\int d^3x \int d^3y \langle {\cal J}^\lambda(x) {\cal
J}^\mu(y)\rangle h_\lambda(x) h_\mu(y)/2~.
\label{SC}
\end{equation}
In the Coulomb gauge, 
the Fourier transform of Eq. (\ref{SC}) can be expressed as
\begin{eqnarray}
S_C =
{ i\sum_{k,\mu} j_\mu(k)j_\mu(-k)\langle {\cal
J}_\mu(k) {\cal J}_\mu(-k)\rangle /(2{\bf k}^2) }~. \nonumber
\end{eqnarray}


\begin{references}
\bibitem{comment1}The resulting quasiparticle fermions would be
multi-valued upon encircling a quantum of  vortex\cite{SIT1}, 
but we ignore such
effects since we do not discuss the single-particle properties.
\end{references}
\end{document}